# Multiscale ultrafast laser texturing of marble for reduced surface wetting

Rocío Ariza[1, 3], Miguel Álvarez-Alegría[1], Gloria Costas[2], Leo Tribaldo[2], Agustín R. González Elipe[4], Jan Siegel[1], Javier Solis[1]

1. Laser Processing Group, Instituto de Óptica, IO-CSIC, Serrano 121, 28006 Madrid, Spain
2. Levantina y asociados de minerales, Autovía Madrid-Alicante s/n, 03660 Novelda (Alicante), Spain
3. Department of Materials Physics, Faculty of Physics, Complutense University of Madrid, Madrid 28040, Spain
4. Nanotechnology on Surfaces Group, Instituto de Ciencia de Materiales de Sevilla (ICMS-US-CSIC), Américo Vespucio 49, 41092 Seville, Spain

## Abstract

The modification of the wetting properties of marble surfaces upon multi-scale texturing induced by ultrafast laser processing (340 fs pulse duration, 1030 nm wavelength) has been investigated with the aim of evaluating its potential for surface protection. The contact angle (CA) of a water drop placed on the surface was used to assess the wettability of the processed areas. Although the surfaces are initially hydrophilic upon laser treatment, after a few days they develop a strong hydrophobic behavior. Marble surfaces have been irradiated with different scan line separations to elucidate the relative roles of multi-scale roughness (nano- and micro-texture) and chemical changes at the surface. The time evolution of the contact angle has been then monitored up to 11 months after treatment. A short and a long-term evolution, associated to the combined effect of multi-scale roughness and the attachment of chemical species at the surface over the time, have been observed. XPS and ATR measurements are consistent with the progressive hydroxylation of the laser treated surfaces although the additional contribution of hydrocarbon adsorbates to the wettability evolution cannot be ruled-out. The robustness of the results has been tested by CA measurements after cleaning in different conditions with very positive results.

## 1. Introduction

Marble is a natural material that has widely been used throughout history and up to the present day, where it holds a prominent position in different market sectors related to architecture and decoration, among others. Yet, its durability is harmed by atmospheric and local acidic conditions and its aesthetic appearance endangered by human pollution and contamination. Protection strategies for marble are mostly based on the improvement of the wetting properties of the surface to prevent staining and acidic chemical attack, as hydrophobicity reduces the contact area of the surface with the degrading agent.

Several works have reported the formation of hydrophobic surfaces in natural stone by the use of chemical coatings [1–3] aiming at minimising the surface energy. Nevertheless, the use of coatings often involves problems associated to adhesion or durability [3], crack formation or yellowing of the coating [4]. Surface structuring by femtosecond laser processing has been reported as an effective way to modify the wettability properties of different materials like polymers [5], dielectrics [6], semiconductors [7] or metals [8–11]. In this respect, it must be emphasized that both, surface topography and surface chemistry influence wettability [12]. Indeed, the wettability in metals after femtosecond irradiation has been widely investigated and a short-to-long temporal evolution has been observed in most of the cases. Kietzig et al. and Bizi-Bandoki et al. explained the time evolution of wettability through the attachment of chemical functional groups at the surface over the time [9,10]. It is remarkable that the generation of superhydrophobicity may be caused by a single monolayer [13]. In general, such chemical layers generated by surface contamination may experience similar problems as artificial coatings that are designed for modifying the wettability of the surface. They can be in addition, not reproducible unless a controlled environment is used during "aging" [14].

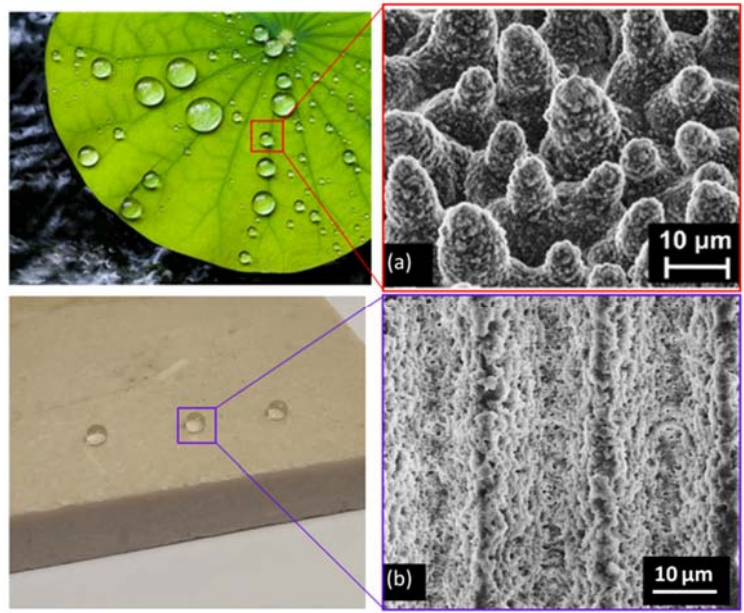

*Figure 1. Hydrophobic behaviour in the lotus leaf (a) and processed marble (b) and their microstructure.*

In general, very little research has been reported on the use of laser processing of natural stone for protection applications [15,16], coatings being the preferred option. These works report on the potential of this technique to achieve hydrophobic surfaces in natural stone, but details regarding the stability and temporal evolution of the wetting properties of the laser treated surfaces were not provided. In this work we aim at analysing the potential for improving the wetting properties of marble through a laser processing strategy based on multiscale (micro- and submicro-meter) texturing of the surface. Such multi-scale texturing process forms the basis of several biomimetic laser processing approaches that have been studied with different goals (i.e., improvement of optical, mechanical or wetting properties) in a huge variety of materials [17]. For instance, natural hydrophobic surfaces have been observed in certain plants like in leafs of the lotus flower, and its replication on silicon and metals by means of laser processing was an early breakthrough in this field [18,19]. It is worth emphasizing again that also in nature, both, multi-scale roughness and surface chemistry, contribute to the surface wettability: for instance, lotus leaf surfaces feature a multiscale topography and are covered by epicuticular waxes [20].

In the context of natural stone protection, most of the harmful substances for the surface quality and aesthetic appearance are water-based liquids or water soluble acidic/basic substances.

Therefore, the improvement of water repellence properties of natural marble is clearly a desirable goal. We have used ultrashort laser pulses to structure the surface and induce a multi-scale type of texture [10,19]. A broad range of processing parameters have been analysed in order to obtain a bioinspired functional modification of the surface morphology that resembles that of a lotus leaf. The latter features a cone structure at the meso-scale (tens of micrometers) superimposed with a surface roughness at the nanoscale, as shown in Figure 1a. The laser-fabricated morphology in marble (shown in Fig. 1b) with an equivalent functional performance of water repellence is composed of trench structures with tens of microns spacing and micrometer depth superimposed by nanoscale roughness. These laser-induced morphologies increase the contact angle (CA) compared to that of pristine surface. The wettability shows in addition a temporal evolution, improving its contact angle over several days, reaching the hydrophobicity regime (CA > 90°). The influence of the induced topography and the chemical evolution of the surface after laser processing have been analysed, as well as the robustness of the achieved results. Overall, the results show that laser processing provides a potential strategy for the wetting control of marble. It must be emphasized, though, that it is the combined effect of laser irradiation and aging (hydroxylation of the laser treated surface and the likely incorporation of hydrocarbon species) that finally conveys the surface with its final hydrophobic behavior after a few days.

## 2. Materials and Methods

The study has been carried out in *Crema-Marfil Coto* marble samples from the Pinoso quarry (Alicante, Spain), supplied by Levantina y Asociados de Minerales S.A. The samples were cut in 5 cm x 5 cm x 1 cm slabs and mirror polished. Petrographic examination was done at Centro Tecnológico del Mármol (CTM) following the European standard UNE-EN 12407:2007. The analysis is based on petrographic optical microscopy of thin sections samples in transmission with parallel and crossed polarizations.

Surface morphology characterization was done, before and after laser processing, by scanning electron microscopy (SEM) in a FEI Inspect working at 15 kV and using a 10-15 mm working distance. The element composition of the samples was analyzed by Energy–Dispersive X-ray spectroscopy (EDX) in a Hitachi TM3000 SEM at 15 kV. The compositional data obtained by EDX corresponds to the material composition averaged over the electron penetration depth (a few microns) and is useful to assess chemical modifications in a thin subsurface layer (bulk) after laser processing [21].

Luminescence properties were investigated by means of cathodoluminescence in a Hitachi S2500 SEM by recording spectra in the visible range with a Hamamatsu PMA-11 charge coupled device camera and working at 15 kV excitation. No discharge-coating was needed for none of the SEM techniques used. The trench depth measurements were obtained using a high NA optical microscope Nikon Eclipse with a high-precision axial position stage.

The laser processing of the marble was carried out using a *Satsuma* HP$^2$ Yb fiber laser from *Amplitude Systemes*. The laser system delivered 350 fs pulses at a wavelength of 1030 nm and a repetition rate in the 1-500 kHz interval. The laser beam polarization was linear and oriented by using a λ/2 waveplate to lie parallel with respect to the scan direction.

The laser beam is deflected over the surface under treatment by a galvanometric mirror scanner (*ScanCube 14*) and focused by F-theta lens with focal length of 100 mm, allowing to scan the laser beam over a large field (7 x 7 cm$^2$) at speeds up to 7 m/s. A telescope and a 5 mm pinhole are placed in the optical beam path to reduce the laser beam diameter at the entrance of the scanner in order to match the entrance pupil of the latter and thus maximize the scanner throughput. The waist radio (1/e$^2$) of the laser beam at the surface was measured leading to a value of 19.4 µm. Fluences per pulse in the 1-4 J/cm$^2$ range were used during the laser processing experiments. The scanner and the laser system are synchronized by means of the *LaserDesk* Software, which also allows performing user-designed irradiation patterns.

The samples were cleaned with distilled water to remove the residual powder and dirt at the surface before laser processing. Once the surface was processed, it was covered by a film of compacted powder caused by the material ablation. A cleaning protocol has been implemented and was followed for all samples. After irradiation, the samples were cleaned with a mixture of soft liquid soap and distilled water in an ultrasonic bath for 10 minutes to remove the powder. Then, they were dried in a muffle furnace at 50°C for 48h, based on a previous study aimed at determining the optimal temperature and time to achieve proper dehydration of the marble. The processed samples were then stored in controlled atmospheric conditions (22°C, RH 30%, air atmosphere) over all the experiment time (>11 months) in desiccator chamber.

To analyze the wettability of the samples, contact angles [22] of distilled water drops placed on fresh regions of the surface were measured at increasing times after processing. A different and "fresh" processed region was always measured in order to assess the evolution of the untreated and laser-treated surfaces without affecting the temporal evolution of surface chemistry. The indicated "Day 0" values correspond to the contact angle recorded just after the cleaning protocol (after drying in the muffle furnace). The measurements have been done with a home-built setup, composed of a micro-dispenser that allows the deposition of droplets with a precise volume, ranging from 5 µl to 50 µl. In all the experiments, the droplet volume was fixed at 5 µl. The sample was illuminated from behind with a white light source and the images were recorded with a CCD camera (*PCO*) 5 seconds after placing the water drop. The contact angle of the droplet with the surface was determined with the "Contact angle" plug-in for the Fiji software. The CA values provided were determined by averaging the results of several measurements. That way, mean values with a standard deviation of typically 5° - 12° were obtained. Also, in some cases (for instance the values measured after aging for one or more months) the determined CA values were obtained following the industrial standard EN 15802:2009 at the facilities of the Centro Tecnológico del Mármol (https://ctmarmol.es/), providing values consistent with the ones measured at the home build setup.

Surface chemical analysis was performed by means x-ray photoelectron spectroscopy (XPS) and Fourier-Transform Infrared Spectroscopy in the range of 400 – 4000 cm$^{-1}$, using the attenuated total reflectance configuration (IR-ATR). XPS measurements were conducted using a VG Microtech MT 5000 spectrophotometer with a non-monochromatic Mg Kα x-ray source operating at 300 W. The pressure in the analysis chamber was maintained below 10$^{-9}$ torr. The energy scale was calibrated with an Ag 3d5/2 (368.3 eV) standard and the peaks have been fitted by Gaussian-Lorentzian mixed functions, after a Shirley background subtraction. The binding energies were calibrated by means of the C1s band at 285.0 eV (adventitious C). The FTIR measurements were carried out in a Perkin Elmer Spectrum 100 system. Samples with different times of aging were analyzed with both techniques in order to investigate the chemical changes over time. The XPS measurements were conducted in Centro Nacional de Investigaciones Metalúrgicas (CENIM-CSIC) and the FTIR-ATR spectra were recorded at the Instituto de Cerámica y Vidrio (ICV-CSIC), respectively.

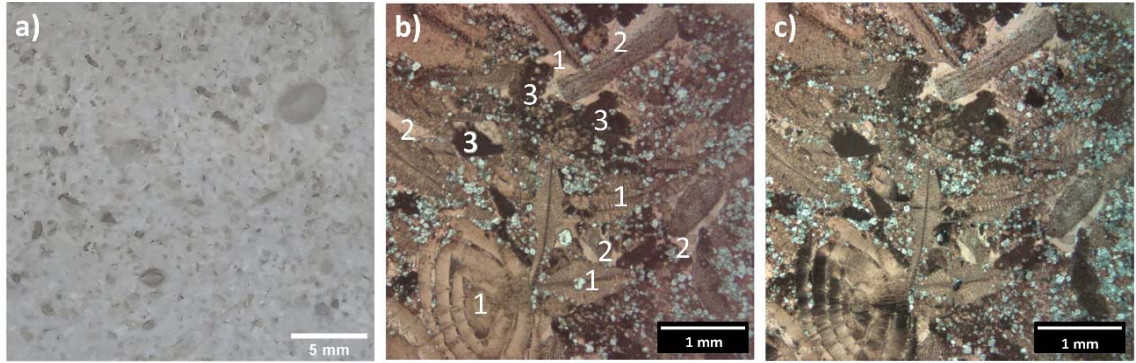

*Figure 2. Visual aspect of Crema-Marfil marble (a) and petrographic microscopy images with parallel (b) and perpendicular (c) polarization. Several representative structures are identified with numbers in Fig.2(b). (1) bioclasts, (2) sparry calcite (white contrasted crystallization features at the edges of bioclasts), (3) micrite (matrix material).*

## 3. Characterization of the pristine material

The samples used in this study were classified as biosparite (Folk classification) or wackestone-packstone (Dunham classification). Its macroscopic aspect shows a light brown appearance where darker veins can be observed. Images recorded with a petrographic microscope (Figure 2) show a carbonate sedimentary rock composed of a wide number of structures with different grain sizes: bioclasts (0.1 to 10 mm), sparry calcite (0.05-1 mm), micrite (matrix) and in some cases dolomite (0.01-0.2 mm). The heterogeneity of the sample and the unique properties of each grain are a challenge for the laser processing.

Limestone is a heterogeneous mixture of calcite ($CaCO_3$) and different amounts of dolomite ($CaMg(CO_3)_2$) as main constituents. The relative concentration of both minerals is a unique distinguishing mark of each marble. This fact has been used over the last decades in archaeology and cultural heritage to determine the quarry origin of marble in archaeological sites [23]. One of the most prominent characterization techniques is cathodoluminescence, which is a non-destructive analysis being able to provide accurate information and aid to identify the quarry of origin [23,24]. Cathodoluminescence (CL) spectra were recorded in this study to assess the starting material and the content of calcite and dolomite in the sample. The CL spectrum in Figure 3a shows two well-defined bands in the UV (intrinsic centres) and orange-red visible (extrinsic centres) range. The first one is related to lattice imperfections and second one is produced by impurities. The presence of impurities in the material can produce an intense luminescence emission [24]. Different ions can easily substitute $Ca^{2+}$ or $Mg^{2+}$ in the lattices, with the most common impurity being $Mn^{2+}$. CL is sensitive not only to the presence of element traces but also it provides relevant information of the lattice position where the impurity is incorporated. Indeed, $Mn^{2+}$ impurity shows different emission bands when replacing $Ca^{2+}$ from $CaCO_3$ (calcite) or from $CaMg(CO_3)_2$ (dolomite). Literature works [23–26] correlate the emission of $Mn^{2+}$ at 620 nm with impurities in calcite, while its emission when incorporated in dolomite is observed at 575 nm and 650 nm. In addition, UV bands have been reported at 355 nm, 396 nm and 475 nm [26].

Figure 3**Error! Reference source not found.**a shows a strong and clear band centred at 620 nm which identifies our sample as calcite type. Due to the spatial inhomogeneity and the presence of darker veins, a series of local measurements were also performed. A few of those spectra showed a peak at 655 nm indicating a low percentage of Mg in the lattice, consistent with the petrographic study. Some spectra with a dominant UV band were also observed. This could be linked to lattice imperfections in calcite [26]. These observations can be correlated to a certain density of defects that can affect the laser absorption, locally changing the ablation threshold.

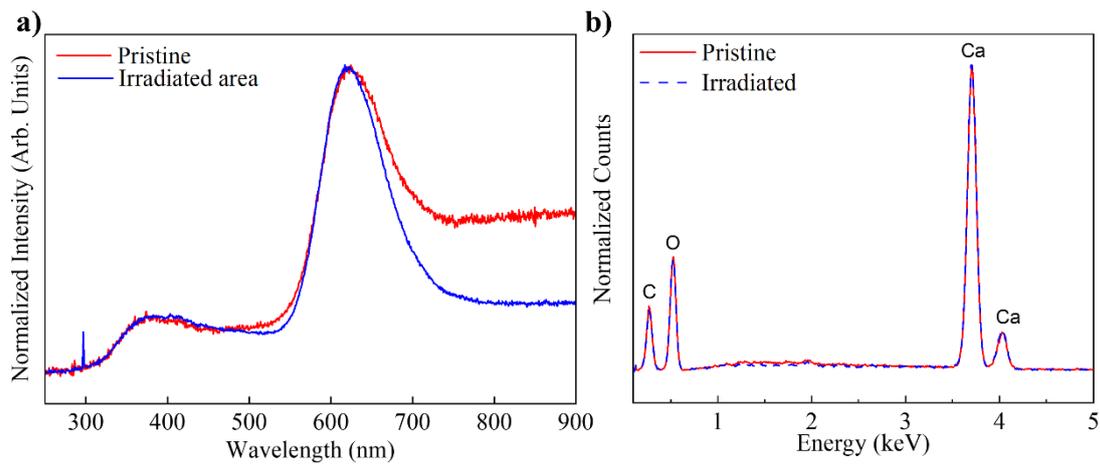

*Figure 3. Processed sample with 20 µm of scan separation. a) Cathodoluminescence spectra and b) energy-dispersive X-ray spectra in laser processed and unprocessed areas.*

The identification of the samples as constituted mainly by calcite is also supported by the EDX analysis. This technique provides information of the chemical elements presents in the sample. As show in **Error! Reference source not found.**b, carbon, oxygen and calcium were detected, before and after laser processing while magnesium traces (peaks at 1.254 and 1.302 keV) could not be identified due to their very low concentration. As shown in Figure 3b, the sub-surface composition of the material is apparently preserved upon laser processing.

## 4. Laser processing effects

### 4.1. Morphology

Many works have reported the formation of laser induced periodic surfaces structures (LIPSS) upon short and ultrashort laser irradiation as a versatile way to modify and to improve certain properties of surfaces. In particular, the effect of LIPSS along with surface chemistry has been used to strongly modify the wettability of different materials [5,7,8,11]. The challenge in natural stone samples is their heterogeneity that causes different interaction mechanisms of the laser pulses with each type of grain, leading to irregular surface absorption. We decided to perform the laser texturing process by exploring different parameters aimed at the fabrication of a texture (meso- and microscopic) similar to that observed in lotus leafs. A detailed analysis of the influence of the different laser processing parameters has been done, including laser power, spot diameter, scanning speed, separation between scan lines and number of scans in order to achieve the best conditions for generating functional textures on marble. Pulse width and frequency were fixed at 350 fs and 500 kHz, respectively, in order to maximize nonlinear absorption and heat accumulation effects, which are both relevant for maximizing energy deposition efficiency [27,28].

As an initial step, the pulse energy required to distinctly mark the surface was determined. This energy threshold was measured in single shot laser exposures. With the focusing optics used, the minimum fluence required to induce surface damage was ~ 3.2 J/cm$^2$. However, in this case, only discrete damage spots at the grains with weak absorption are observed. For fluences above ~ 3.7 J/cm$^2$, most of the treated spots were damaged. If the laser is scanned with a high overlapping factor among consecutive pulses (typically above 90%), the cumulative effect (incubation [29] and heat accumulation [27,28]) of successive pulses leads to a strong decrease of the damage threshold. The damage threshold observed for spatially overlapping pulses was ~ 1.0 J/cm$^2$. Therefore, in the following experiments we have used a pulse fluence of 1.6 J/cm$^2$, well above threshold, to ensure a homogenous irradiation of the whole processed region.

In order to generate a dual scale roughness structure on the surface for the enhanced control of the wettability [10,19] , a scan pattern consisting of parallel lines was designed with a spacing ranging from 2 µm to 100 µm. The scan velocity was fixed at 50 mm/s and a single scan was performed in all cases. Changes in the pulse energy (above the threshold) or scan number do not vary significantly the obtained morphology of the trenches and only slight differences in the depth are appreciated.

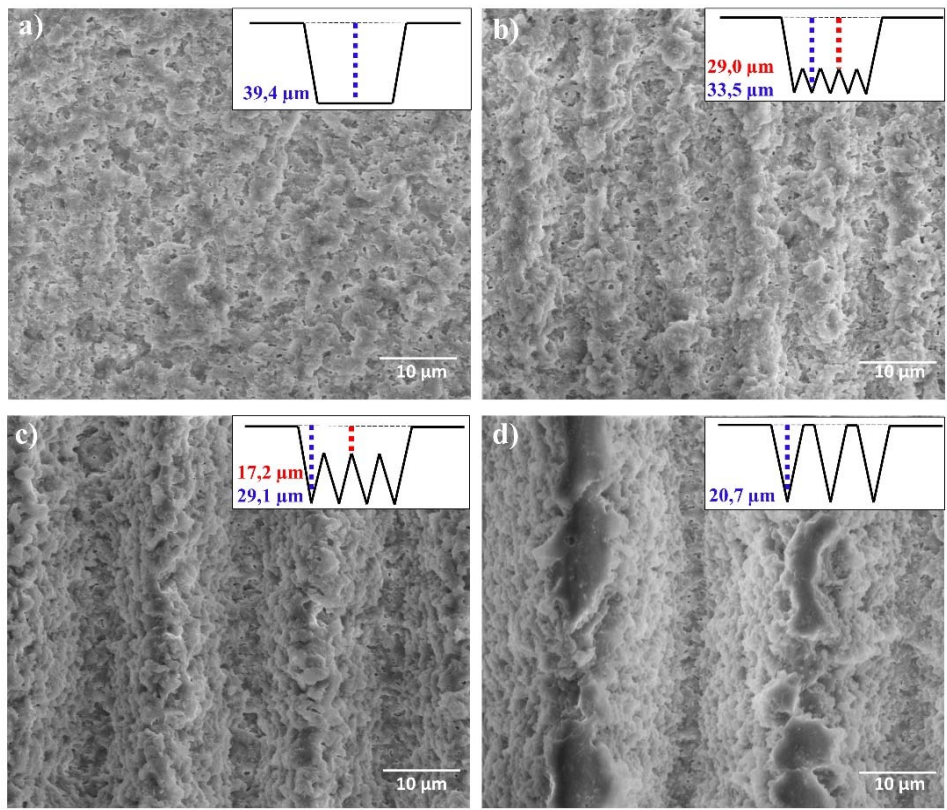

*Figure 4. SEM images of areas with a) 2 µm, b) 10 µm, c) 20 µm and d) 30 µm of the scan line separation, indicated as Δ. The width of the trenches varies depending on the line overlap. The inset shows a sketch of the trenches depth profiles and the measured depth values in each case.*

Figure 4 shows several SEM images corresponding to different scan line separations, illustrative of the evolution of the surface morphology with this parameter. When the scan separation is considerably smaller than the laser spot radius at the surface (19.4 µm), the processed area shows a uniform roughness distribution, without observable trenches. However, for separations from 10 µm onwards vertical trenches are clearly seen. For a separation above 30 µm, the trenches are completely separated by regions of non-modified material. Due to the Gaussian intensity distribution of the laser at the surface, the depth profiles of the trenches depend on the line separation. The insets in Figure 4 show a sketch of the depth profiles of the induced morphologies. In preliminary experiments it was observed that ablated depth mainly depends on the line separation and the energy used, whereas the influence of the scan velocity is minor. We thus decided to keep this parameter as fixed (50 mm/s) as above indicated. The observed role of the laser processing parameters in the marble is similar to that reported in literature for natural stone where the laser pulse energy is the key parameter [15,16]. In these works, the authors provide estimates for the expected topographic changes caused by ablation in the surfaces of granite [15] or marble [16] and, as in our case, do not observe appreciable compositional changes in the subsurface upon laser processing. They correlate the wettability changes at the surface to the induced roughness, without considering its temporal evolution as a consequence of slow evolution changes in the surface energy.

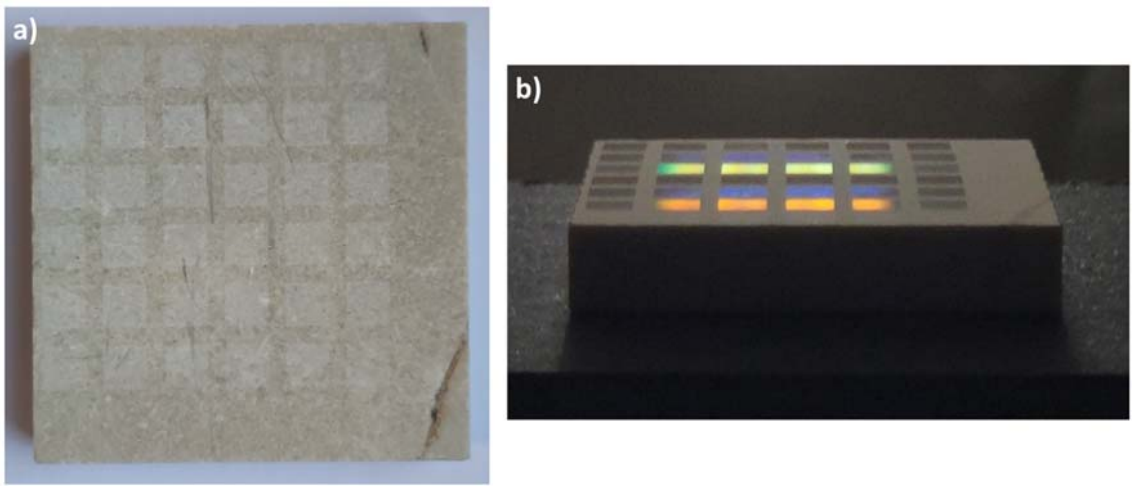

*Figure 5. a) Visual aspect of irradiated areas on the marble sample and b) the light diffraction effects observed upon white light illumination at an angle of 75º from the normal and observation at almost grazing angle.*

As it can be seen in Figure 5a, the visual aspect of the processed surface keeps nearly unaltered, and only a slight bleaching is observed in the processed areas. However, due to the anisotropy of the patterns produced (line arrays), the optical appearance varies under certain illumination conditions: strong light diffraction effects in form of a wavelength-selective diffraction angle (related to the scan line separation) upon white light illumination can be observed, as shown in Figure 5b. In addition, not only the visual aspect remains unmodified, but also the subsurface material composition is preserved upon the laser irradiation as is shown in the CL and EDX spectra in Figure 3**Error! Reference source not found.**. These results demonstrate that the processing scheme used does not lead to substantial changes, neither in the aesthetic appearance of the surface nor in the subsurface composition of the material.

## 4.2. Wettability and temporal evolution

The wettability of a material surface is a complex property where multiple other properties are involved. It is known that the wettability depends on the surface roughness and the surface free energy, which is why it is possible to modify the contact angle varying both of these parameters [30]. The surface free energy is determined by the chemical composition, therefore covering the surface with specific chemical functional groups may reduce the surface energy and increase the contact angle as predicted by Young's equation [30,31]. The most hydrophobic functional groups are based on fluorine due to its large electronegativity and covalent bond with carbon (i.e - $CF_3$ < - $CF_2H$ < - $CF_2$ < - $CH_3$ < - $CH_2^-$) [32]. On the other hand, functional groups such as C=O, - $NH_2$, -COOH, -$OSO_3H$… enhance hydrophilicity [5,30].

Morphology and roughness, nevertheless, also affect the wettability of a surface. There are two basic models to explain the role of roughness on the wettability. Wenzel's model [33] considers that the liquid droplet fills up completely the topography of the surface and enhances the intrinsic wetting properties of the material: more hydrophilicity in hydrophilic materials and more hydrophobicity in hydrophobic materials. However, Wenzel's model is limited to small roughness values. When roughness is increased the Wenzel`s equation fails and show a $\cos\theta$ exceeding 1 [34]. Beyond this limit, Cassie-Baxter's model [35] considers the presence of air trapped inside the topographical features and predicts always a change towards hydrophobic behaviour.

Laser processing induces several changes at the marble surface: It creates a pronounced multi-scale roughness and "initializes" the surface in terms of chemical reactivity. The contact angle measurements bring thus significant information regarding the roles roughness and chemical evolution of the surface on the wettability.

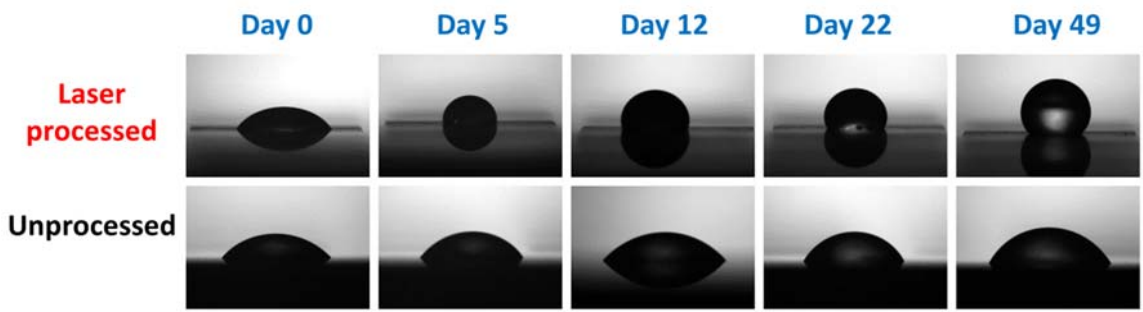

*Figure 6. Back-illumination images of water droplets on a laser processed region (top row) and a pristine surface (bottom row) for different times after irradiation. These images enable measuring the evolution of the contact angle (CA) vs. time. The laser-processed surface corresponds to a scan line separation of Δ = 2 µm, representative of the global behaviour of the laser treated material.*

The changes in the wettability properties of the treated surfaces were monitored by means of static contact angle measurements using back-illumination images of water droplets deposited on the surface, as shown in Figure 6. The images allow measuring very accurately the contact angle between the surface and the droplet. It must be emphasized, though, that in many cases, the accuracy of the CA measurements is also conditioned by the presence of local minima in the Gibbs free energy of the liquid on a rough surface [22]. These measurements have been repeated over time, always on dry areas to avoid the influence of trapped $H_2O$ molecules at the surface, in order to assess the evolution of untreated and laser-treated surfaces. Figure 6 shows a representative sequence of the droplet behaviour and its temporal evolution for both cases. "Day 0" refers to measurements made immediately after the cleaning protocol and drying. The "Day 0" values should thus correspond to the wettability changes associated to topographic (roughness) effects not yet conditioned by the incorporation of chemical functional groups at the surface over time.

The time evolution of the CA for four representative scan line separations is shown in Figure 7, along with measurements performed in non-processed regions of the same samples. We can distinguish two different types of evolution depending on the scanline separation, leading to the formation of trenches (micro-scale roughness) or uniformly rough regions (strong overlap between adjacent scans, that supresses the micro-scale roughness). This difference is not surprising since the roughness is important to maximize the effective area, understood as the total surface area that contributes to the interaction with a water drop. Indeed, Gao et al. [36] studied the importance of two length roughness scales in the "Lotus effect". In our case, scan separations from 10 µm and higher lead to the formation of trenches, creating a micro-scale roughness. The surface of the trenches is covered by flake-like structures showing a nano-roughness.

Figure 7a shows the temporal evolution in the area patterned with 2 µm of scan separation. The morphology of this area displays only the nano-roughness from flakes since trenches were not created due to high overlapping. Immediately after laser treatment, the processed area is more hydrophilic than the unprocessed one, which might suggest that its behaviour is governed by the Wenzel's model. However, the CA shows then a rapid increase with time over the following days, reaching a value above that of the non-processed surface within a few days. This increase continues for longer times, although at a slower rate. The highest CA recorded was 140°, after 340 days (11 months) of aging. This result is important for understanding the relative roles of micro-and nano-roughness in the wettability behaviour. In absence of micro-roughness, laser texturing by itself does not improve the wettability immediately after processing. However, the irradiation and the formation of the nano-scale roughness activates and enhances chemical changes at the surface that are responsible for the long-term evolution of the wettability. The increase of the effective surface area of the material due to the nanoscale roughness may be a key factor behind the larger surface reactivity with functional groups present in the environment.

When the line separation is increased the trenches split and micro-roughness appears. Figure 7 b, c, d show the temporal evolution of the areas where both nano- and micro-roughness are induced. There are two remarkable aspects that have to be considered. The first one is that the line pattern induces a noticeable anisotropy in the droplet shape and two different contact angles are observed, depending on whether the sample is oriented with the trenches parallel or perpendicular to the imaging axis. This phenomena has been observed in previous works on other materials [8,11,37] and results in a consistently higher CA for measurements parallel to the lines. The second is that, in all cases, the CA obtained from measurements parallel to the trenches is always above the value of the pristine surface immediately after processing. Such "Day 0 measurement" provides thus practical information of the CA improvement exclusively due to micro-roughness.

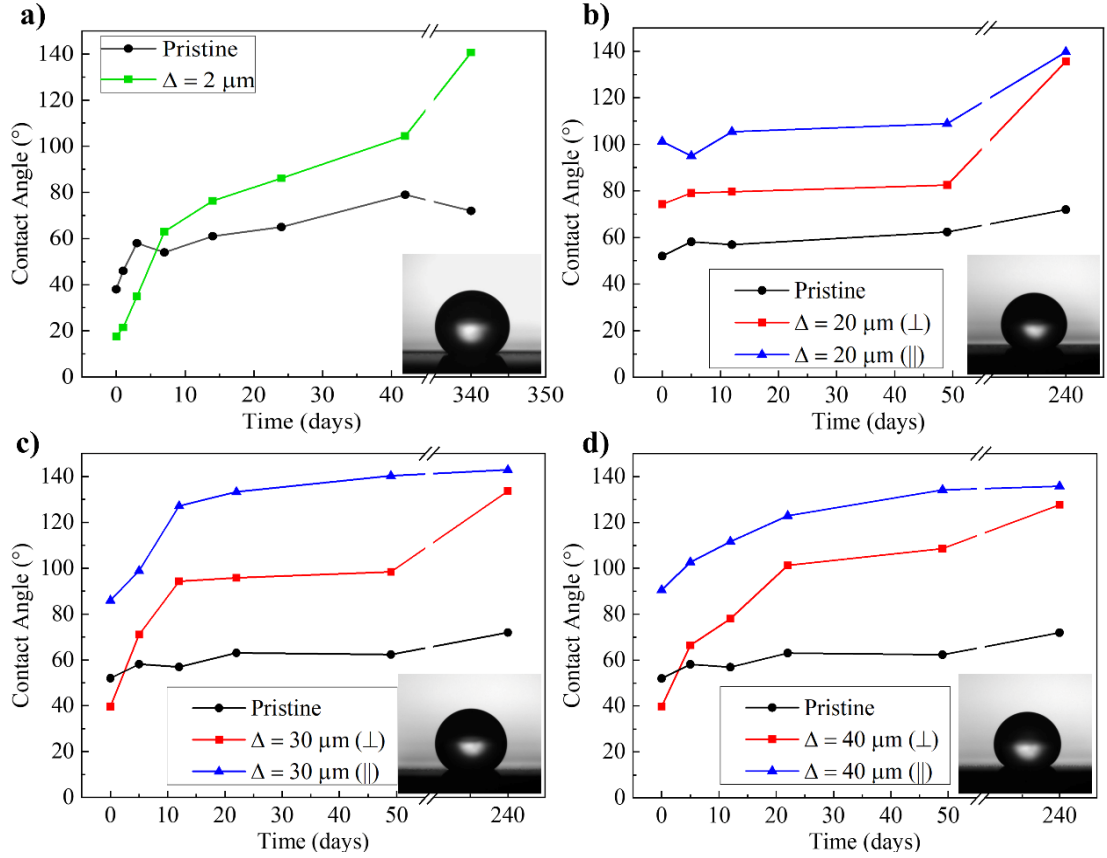

*Figure 7. Temporal evolution of the contact angle for samples processed with a scan line separation of a) Δ = 2 μm, b) Δ = 20 μm, c) Δ = 30 μm and d) Δ = 40 μm. For b)-d), the CAs show a noticeable anisotropy due to the line pattern and CA measurements for the trenches oriented perpendicular and parallel to the imaging axis were recorded. The inset shows the drop image for the parallel measurement made on the last day of recorded data series. Standard deviation of the CA measurements ranges typically from 5º to 12º.*

In contrast, the CA at "Day 0" from the measurements perpendicular to the trenches is above that of the pristine surface only in the sample processed with a scan line separation of Δ = 20 μm. Interestingly, for higher separations (i.e 30 μm, 40 μm), the time evolution of the CA perpendicular to the trenches strongly resembles the evolution of the sample with Δ = 2 μm. The time evolution of the perpendicular CA is weaker for Δ = 20 μm, which can be related to the increase of effective surface as the scan line separation is increased, also evident in the sketches in Fig.4. Although the highest CA is obtained from the parallel measurements, both types of measurements tend to converge over time. The CA values obtained in the saturation regime are similar to those reported in literature in granite [15] and are close to a super-hydrophobic behaviour.

The micro-roughness was adjusted to Wenzel model and Cassi-Baxter model and compared with the experimental results in Table 1. The equations applied for each mode for the trench geometry are [38]:

$$Wenzel: \cos\theta^W = \left[1 + \frac{2h}{a+b}\right]\cos\theta^{Surface}$$

$$Cassi-Baxter: \cos\theta^{CB} = \left[\frac{a}{a+b}\right]\left(\cos\theta^{Surface} + 1\right) - 1$$

where *a* is the peak width, *b* is the trenches width, h is the trench depth and *θ* the contact angle.

|  | Cassie-Baxter model | Experimental |
|---|---|---|
| Δ = 20 µm | 101,1° | 101° ± 8° |
| Δ = 30 µm | 101,1° | 86° ± 6° |
| Δ = 40 µm | 89,4° | 90° ± 5° |

*Table 1. Contact angles obtained for the micro-roughness induced in the surface with separation of Δ = 20, 30 and 40 µm by Cassie-Baxter model and the experimental results. Wenzel model is not shown due to the high roughness induces a failure in the model as expected.*

The micro-roughness is larger than the Wenzel model can support and, therefore, topography cannot be adjusted by this model. However, the Cassie-Baxter model results approach, especially for Δ=20 µm and 40 µm the measured experimental values "at Day 0". The best fitted CA corresponds to Δ = 20 µm, showing this trench separation provides results consistent with the Cassie-Baxter model.

The results in Figure 7 clearly show that trenches with Δ = 20 µm produce the maximum initial improvement related to laser induced micro-roughness alone. This trench separation maximises the behaviour predicted by the Cassie-Baxter's model. In this model, the aspect ratio and the separation is crucial, just as convex structures with smooth edges [39,40]. However, the higher parallel CA values achieved for Δ = 30 µm and 40 µm for a relatively short aging (~ 130° in ~20 days), indicate that the increase of effective surface plays an essential role to enhance the wettability properties by chemical changes at the surface. For the three separations, along with the production of the micro-roughness, the initialization of the chemical state of the surface leads to a long-term evolution associated to the incorporation of functional groups at the surface. For this mechanism, the best results are observed when the effective surface covered by functional groups is larger (i.e. Δ = 30 µm).

### 4.3. Surface chemistry assessment and temporal evolution

Regarding the specific chemical mechanisms responsible for the temporal evolution of the wettability, it must emphasized that the evolution of the CA over time is a well-known behaviour in laser processed metals [9–11]. For instance, laser interaction with metals results in a progressive attachment of molecules to the surface. Kietzig et al. [10] observed an increase of carbon in the steel surface as a function of time, up to several days after processing. Several functional groups containing carbon are non-polar, and their attachment to the surface may reduce the surface energy. In contrast, Bizi-Bandoki et al.[9] observed the opposite effect: reduction of carbon amount in the steel surface. They attributed the observed hydrophobicity increase in the processed surfaces to the removal of $H_2O$ molecules from the surface, making the irradiated surface less polar and therefore improving slightly its hydrophobicity. In the case of Al, the authors observed an increase in the non-polar C-content of the surface that they correlated to the formation of pyrolytic C and the attachment of –CH3 groups resulting in surface hydrophobicity.

Although similar processes could be expected to occur at the processed marble surface (a relatively porous material), the temporal evolution of the CA reported in Ref. [8] covers just a few days, and we have therefore not enough grounds to ascribe the behaviour we observe to this single mechanism. Alternatively, we can also consider the role of surface hydroxylation [41].

Several works have correlated the formation of hydroxyl groups to the evolution of the wettability of metal oxides [42] including laser processed surfaces [43]. In our case, we have to consider also the possibility that pristine $CaCO_3$, which is hydrophilic [44], can be thermally decomposed at relatively low temperatures. Indeed, the thermal conversion of calcite to calcium oxide (CaO) is initiated at a slow rate and occurs very rapidly beyond 750°C [45], a temperature easily overcome upon fs-laser treatment. Still, CaO is similarly hydrophilic [46], but reacts with moisture in the atmosphere to form $Ca(OH)_2$, even at room temperature. The longer term temporal evolution observed for the wettability might be thus related to the progressive incorporation of hydroxyl groups to the surface. This, as discussed by Gentleman [41], may lead to a progressive decrease of the wettability of the surface over time (CA increase), more pronounced in processed samples with a large effective surface. This should be considered as a plausible explanation since the relative roles of carbon organic groups and hydroxyls at the surface in the long-term time evolution of wettability upon fs-laser is still under debate [47]. The modification of the near-surface chemical properties does not involve appreciable changes in the "bulk" composition of the material after laser processing. This is evident in the EDX spectra in Figure 3, where the material, probed by the SEM electron beam to a depth of a few microns, shows the same composition before and after processing.

As described in detail in Section 2, the assessment of the chemical changes occurring in the irradiated surface and their temporal evolution has been performed by analysing pristine and laser treated samples by XPS and infrared attenuated total reflectance (IR-ATR) spectroscopies.

Figure 8 shows XPS spectra of the corresponding to the O1s, C1s and C2p bands of a pristine surface and a processed one using $\Delta$= 2 µm scan separation followed by aging for 11 months ($\Delta$ = 2 µm - Aged). The bands have been de-convoluted to analyse the contribution of different species. In the case of the O1s band, it is possible to distinguish two main contributions. At $\sim$ 531.5 eV, a very intense band, corresponding mainly to $CO^{2-}_3$ [48], can be observed. This band may also include contributions from other species with very similar binding energies, like $OH^-$, $O^-$, $O^{2-}_2$. An additional, much weaker band at $\sim$ 533 eV can be ascribed to the contribution of adsorbed $H_2O$ [49]. These two bands are observable both in the pristine and laser processed sample ($\Delta$ = 2 µm - Aged). In literature, a lower energy band located at $\sim$ 529.4 eV has been ascribed to oxide species [49]. This band is negligible in our case for both samples analyzed. The C1s band is formed by three main contributions, at $\sim$ 285 eV, $\sim$ 287 eV and $\sim$ 289.5 eV that are related respectively to C-C and C-H bonds (including adventitious C species), C=O bonds, and carbonate species [48]. Finally, de Ca2p bands show the contributions of the Ca2p1/2 and Ca2p3/2 states located respectively at $\sim$ 351 eV and $\sim$ 347 eV.

A comparison of the different spectra in Figure 8 shows only two easily distinguishable differences in the behavior of the both samples (pristine and laser treated & aged). The first one is the increase of the C1s ($CO^{2-}_3$) band at 289.5 eV in the laser processed sample. This increase is accompanied a by a decrease in the contribution of the C=O band at 287 eV with respect to the pristine sample.

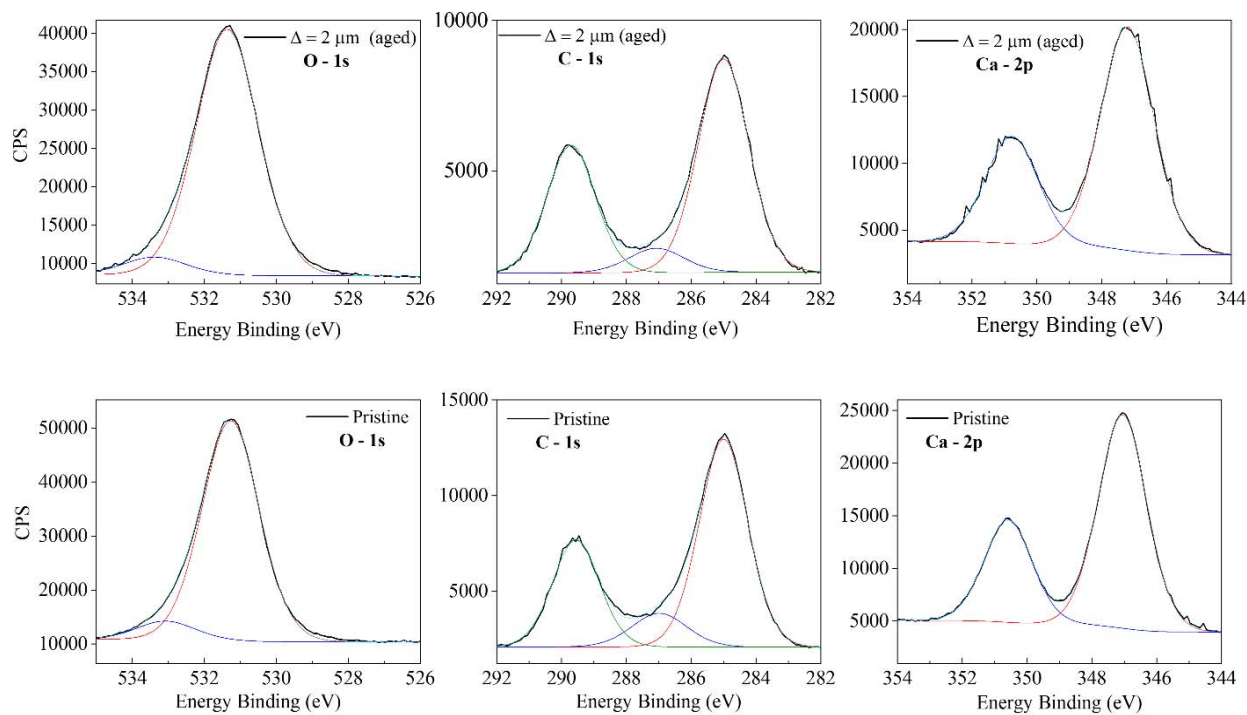

*Figure 8. XPS spectra of O1s, C1s and Ca2p bands recorded at the surface of a pristine sample and a laser processed sample with Δ = 2 μm scan separation after 11 months aging (d2 – Aged). The spectra have been deconvoluted to analyse the contribution of different chemical species to the XPS signal, as described in the text.*

More information can be obtained from the quantification of the surface composition of the two samples shown in Table 2. It can be seen that there is a substantial increase of the oxygen content at the surface of the aged, laser-processed sample that could be consistent with the formation of $OH^-$ species (surface hydroxylation). This increase is accompanied by a diminution in the percentage of carbon at the surface. Interestingly, the Ca2p/C1s (289 eV) ratio, associated to the carbon in the form of $CaCO_3$, is closer to unity in the laser processed sample than in the pristine one. This could suggest that laser irradiation removes part of the carbon species attached to the surface in the form of C=O, C-H and C-C, likely as a consequence of the polishing process.

| Element or XPS Band | Pristine | d2-Aged |
|---|---|---|
|  | Atomic.% | |
| **O (1s) (total)** | **43.5** | **47.2** |
| ∼ 531.5 eV | 40.1 | 44.1 |
| ∼ 533 eV | 3.4 | 3.1 |
|  |  |  |
| **C (1s)** | **44.9** | **39.4** |
| ∼ 285 eV | 26.7 | 23.2 |
| ∼ 287 eV | 4.9 | 2.9 |
| ∼ 289 eV | 13.3 | 13.3 |
|  |  |  |
| **Ca (2p)** | **11.6** | **13.5** |

*Table 2. Surface chemical composition of a pristine sample and a d2-sample aged for eleven months derived from the XPS spectra of both surfaces. The composition associated to the different XPS bands is indicated.*

The indicated changes in the surface composition after laser processing and aging (increase of surface oxygen and diminution of carbon) are consistent with the scenario above described of surface hydroxylation via thermal decomposition of the calcium carbonate, and progressive formation of $Ca(OH)_2$ by reaction with the environmental humidity. In this respect, it is worth noting that Long and co-workers [47] have correlated the formation of hydroxyls with an enhanced adsorption of organic compounds. They observed 30 days after processing of an Al sample with ps-laser pulses an increase of both Al-OH bonds and C content (especially of the C=O contribution at the surface). Even when in our case we observe a clear diminution of the C-content after laser processing and aging, we have not enough grounds to ascribe the hydrophobicity transition a few days after laser treatment solely to the hydroxylation of the surface. To conclude this part of the analysis, it must be noted that XPS spectra acquired in other laser treated samples with different scan separation show a similar trend: about a week after processing, the relative C content of the surface starts to decrease and the one of O to increase with respect the pristine sample.

In order to gain a further insight in the evolution of the surface chemical composition, part of the samples were also analyzed by IR-ATR. For the identification of the IR bands we have used the discussion in references [50] [51]. The peaks corresponding to dry $CaCO_3$ are indicated in Table 3.

| Vibrational mode assignments $CO_3^{2-}$ | Frequency ($cm^{-1}$) |
|---|---|
| $\nu_2$, out-of-plane bend | 712 [50], 714 [51] |
| $\nu_3$, asymmetric stretch | 874 [50], 878 [51] |
| $\nu_1$, symmetric stretch | 1082 [51]-1087 [50] |
| $\nu_4$, in-plane bend | 1435 [51] |
| fundamental stretching ($\nu_4$) vibration in the bulk (intense broad-band) | 1350-1650 [50] |
| ($\nu_1 + \nu_2$) combination band for bulk calcite | 1795 [50] |
| ($\nu_1 + \nu_3$) combination band for bulk calcite | 2515 [50] |
| ($\nu_1 + \nu_3$) combination band for bulk calcite | 2580 [50] |

Table 3. Assignment of the vibrational modes of carbonate ion in bulk dry $CaCO_3$ according to indicated references.

The IR-ATR spectra corresponding to both samples (pristine and laser treated & aged) is included in Figure 9. The roughness of the laser treated sample induces a noise artifact in those spectral regions where the absorbance is close to zero, as it can be seen, for instance, in the 1800-2500 $cm^{-1}$ spectral region. In order to compare both spectra (showing a very different number of counts) both spectra have been normalized to their peak absorbance value at 878 $cm^{-1}$ and the base line at 3550 $cm^{-1}$ has been subtracted. The two intense peaks at ~714 $cm^{-1}$ and ~ 878 $cm^{-1}$ corresponding to $CO_3^{2-}$ are clearly appreciated in both samples. In the region from ~ 1000 $cm^{-1}$ to 1600 $cm^{-1}$ the pristine (polished sample) shows a behavior similar to the one reported by Neagle et al. in high pressure compressed $CaCO_3$ pellets [50]. In contrast, the laser treated sample shows a spectrum very close to the one reported by Al-Hosney and coworkers [51] in non-compressed powder. The differences in this spectral region thus appear to correspond to the roughness/density differences between the two samples. Other bands beyond 1500 $cm^{-1}$ are visible in both samples, although those around 2500 $cm^{-1}$ are smeared out in the laser treated one by the noise artifact above described.

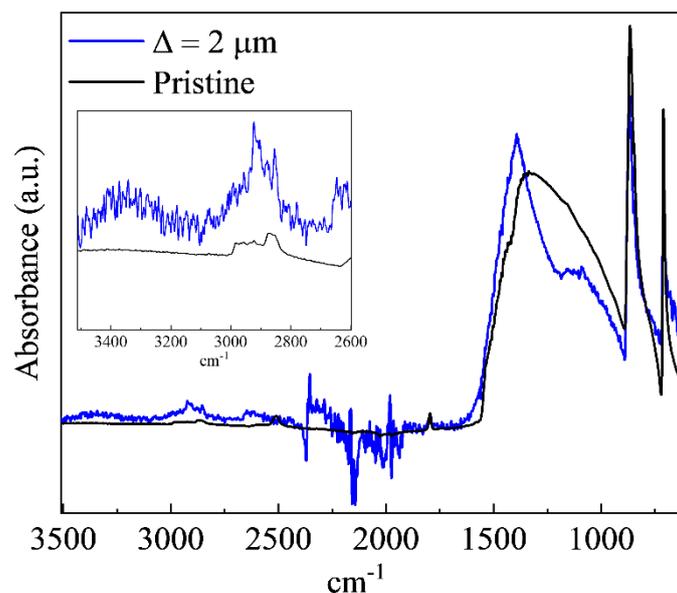

*Figure 9. IR-ATR spectra of a pristine sample and sample d2 O1s, C1s and Ca2p bands recorded at the surface of a pristine sample and a laser processed d2-sample after 11 months aging. The spectra have been deconvoluted to analyse the contribution of different chemical species to the XPS signal, as described in the text.*

For what concerns the presence of hydroxyls and/or hydrocarbon species at the surface, the laser treated sample shows two much weaker broad-bands in the 2600-3500 cm-1 region. The first one is a small band centered around 2900 cm$^{-1}$ that is also present in the pristine sample. In contrast, the second one, centered around 3400 cm$^{-1}$, only appears in the laser processed samples. In order to make these bands more visible, the inset in Fig.9 shows a zoom of this spectral region. The inset unambiguously shows the characteristic OH-stretching band centered around 3400 cm$^{-1}$ [50][51], as well as a band with several peaks that could correspond C-H bonds (C-sp3) around 2900 cm$^{-1}$. The adsorption of hydrocarbon species is also evident in the pristine sample, although it shows hydrophilic behavior. The IR-ATR spectra allow to conclude that the laser treated and aged surface is hydroxylated but shows the additional adsorption of hydrocarbon groups in a larger amount than that observed for the pristine material. While the presence of C-H groups in the pristine material surface does not lead to hydrophobicity, their higher concentration in the aged, laser-treated surface does not allow to ascribe the surface observed hydrophobicity to surface hydroxylation alone.

### 4.4. Durability and robustness of the wettability changes of the surface

The long-term mechanisms responsible for the wettability evolution over time also condition the durability and robustness of the induced modifications. Some functional groups are attached to the surface by Van der Waals forces and their adhesion can be weak. For this reason, is important to monitor the durability of the chemical groups attached at the marble surface. Figure 9 shows the contact angle changes after laser processing, aging and subsequent cleaning of the surface using two different methods. a) Rinsing off the surface with distilled water followed by the removal of liquid droplets by blowing with high purity $N_2$, and drying at 22°C in a desiccator with a 33% RH for 24 h. b) Rubbing the surface with a soft cloth soaked in soapy water followed by the removal of liquid droplets by blowing with high purity $N_2$, and drying at 22°C in a desiccator with a 33% RH for 24 h.

The starting CA for this study was chosen to be after 11 months of aging. The results show that the processed areas keep the contact angle almost unaltered after the first cleaning procedure (distilled water rinsing). For the second process, a significant decrease in the contact angle was observed. This behaviour is consistent with the formation of stable hydroxyls at the surface upon laser processing, which can be partly wiped away in solution applying pressure. The partial elimination of hydrocarbon contamination from the surface could also explain the observed behaviour.

Analysing Figure 10 from the point of view of the previous discussion, two areas with different hydrophobicity origin were compared. As above discussed, the increase in the CA in areas irradiated with $\Delta$ = 2 µm comes mainly from the chemical evolution of the surface. In the other case ($\Delta$ = 10 µm), the CA increase is caused by a combination of micro-roughness and chemical evolution. Taking a detailed look at Figure 8, cleaning with soap produces the maximum reduction in both samples being more noticeable in the area with 2 µm, which suggests that the reduction comes mainly from removing attached molecules from the surface. However, despite the slight reduction, the surface recovers the reference value in a few of days (data not shown). These results demonstrate that the generated hydrophobicity is relatively robust and is up to some extent self-healing.

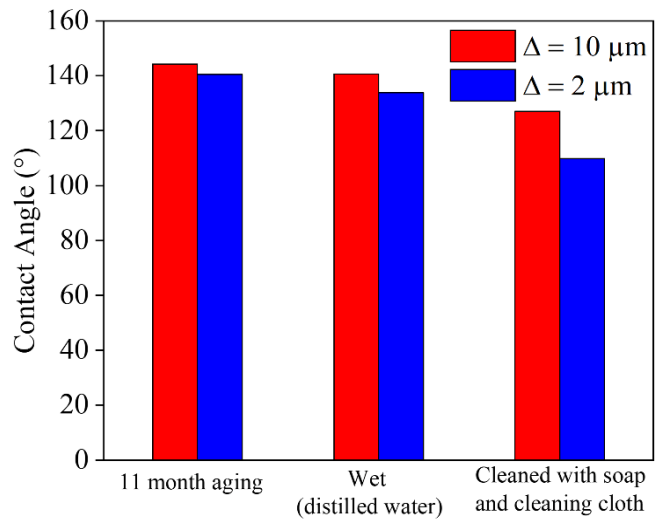

*Figure 10. Contact angle measurements in areas fabricated 11 months ago with Δ=10 µm and Δ=2 µm areas before (left) and after different cleaning processes: soaked in distilled water (middle) and scrubbed with soap and cleaning cloth (right).*

## 5. Conclusions

*Crema-Marfil Coto* marble samples have been homogeneously processed by means of femtosecond laser processing. This technique induces a multi-scale roughness formed by trenches with different separations (micro-roughness) and depths of tens of micrometers, covered by a flake-like morphology (nano-roughness). In addition, the laser treatment triggers a chemical reaction that leads to a noticeable increase of the wettability over the time. All samples show a rapid and substantial rise of the contact angle within the first 10 days. Afterwards, the contact angle increase slows down, stabilizing around 140° at 8 – 11 months of aging. Laser processing thus triggers to two different effects: an "initialization" of the chemical state of the surface that turns much more reactive (likely because the thermal decomposition of $CaCO_3$ in the near surface laser treated region) and an increase in the surface to volume ratio favoring the extent of incorporation of water (hydroxylation) and (likely) C-H contaminants.

The time dependent study allowed disentangling the individual contributions of roughness and chemical evolution at different scan separations. We have found that trenches with a separation of $\Delta$ = 20 µm show the highest initial improvement of the hydrophobicity, attributed to micro-roughness. However, when taking into account the evolution over time associated to chemical reactions, the best results were obtained for $\Delta$ = 30 µm, where the effective surface area is maximized. The mechanisms responsible for the enhanced wettability demonstrated to be robust and durable to cleaning, recovering its reference values in few days after the cleaning process. It can be concluded that laser processing is a useful technique for turning marble surfaces hydrophobic, offering protection to marble surfaces against water-based contaminants without compromising their visual aspect.

## Acknowledgements

This work was financially supported by the EU through project FET-ILP 852048 (BioPromarL), the Spanish Research Agency (MCIU/AEI/Spain) through projects TEC2017-82464-R, RTI2018-097195B-100 and PID2020-112770RB-C21, UCM-Santander 2019 (PR87/19-22613) and the National Research Council of Spain (CSIC) through project 2020AEP049. R. Ariza acknowledges the European Social Fund (ESF) and the Youth Employment Initiative (YEI) of the Madrid region through the PEJD- 2019-PRE/IND-16755 grant. We acknowledge the help and assistance of the FINE-UCM Group (Madrid) for providing access to the SEM-related measurements, as well as the Technical Services of CENIM-CSIC and ICV-CSIC respectively for the XPS and ATR measurements.